\documentclass[a4paper, amsfonts, amssymb, amsmath, reprint, showkeys, nofootinbib, twoside, superscriptaddress, prb]{revtex4-2}

\usepackage[english]{babel}
\usepackage[utf8]{inputenc}
\usepackage[colorinlistoftodos, color=green!40, prependcaption]{todonotes}
\usepackage{amsthm}
\usepackage{mathtools}
\usepackage{physics}
\usepackage{xcolor}
\usepackage{graphicx}
\usepackage[left=23mm,right=13mm,top=35mm,columnsep=15pt]{geometry} 
\usepackage{adjustbox}
\usepackage{placeins}
\usepackage[T1]{fontenc}
\usepackage[pdftex, pdftitle={Article}, pdfauthor={Author}]{hyperref} 
\usepackage{subcaption}
\usepackage{siunitx}
\usepackage{xspace}     
\usepackage{comment}
\newcommand{\BiSe}{Bi$_2$Se$_3$\xspace}            
\newcommand{\subt}[2]{#1$_{\text{#2}}$\xspace}     
\newcommand{\etal}{\textit{et al.}\xspace}         

\begin{document}
\title{Bulk conducting states of intrinsically doped \BiSe }

\author{R. T. Paulino}
    \altaffiliation{Present Address: Department of Physics, McGill University, Montreal, Quebec,   H3A 2T8 Canada.}
    \affiliation{Centro de Ciências Naturais e Humanas, Universidade Federal do ABC, Santo André, SP, 09210-580 Brazil}

\author{M. A. Avila}
    \affiliation{Centro de Ciências Naturais e Humanas, Universidade Federal do ABC, Santo André, SP, 09210-580 Brazil}

\date{\today} 

\begin{abstract}

With a large band gap and a single Dirac cone responsible for the topological surface states, \BiSe is widely regarded as a prototypical 3D topological insulator. Further applications of the bulk material has, however, been hindered by inherent structural defects that donate electrons and make the bulk conductive. Consequently, controlling these defects is of great interest for future technological applications, and while past literature has focused on adding external doping elements to the mixture, a complete study on undoped \BiSe was still lacking. 
In this work, we use the self-flux method to obtain high-quality \BiSe single-crystals in the entire concentration range available on the phase-diagram for the technique. By combining basic structural characterization with measurements of the resistivity, Hall effect and Shubnikov-de Haas (SdH) quantum oscillations, the effects of these impurities on the bulk transport are investigated in samples with electron densities ranging from $10^{17}$ cm$^{-3}$ to $10^{19}$ cm$^{-3}$, from Se-rich to Bi-rich mixtures, respectively, evidencing the transition into a degenerate semiconductor regime. We find that electron-donor impurities, likely Se vacancies, unavoidably shift the Fermi level up to $200$ meV inside the conduction band. Other impurities, like interstitial Bi and Se, are shown to play a significant role as scattering centres, specially at low temperatures and in the decoherence of the SdH oscillations. Previous open questions on \BiSe, such as the upturn in resistivity below $30$ K, the different scattering times in transport and quantum oscillations, and the presence of additional low mobility bands, are addressed. The results outlined here provide a concise picture on the bulk conducting states in flux-grown \BiSe single crystals, enabling better control of the structural defects and electronic properties.

\end{abstract}


\maketitle

\section{Introduction}\label{sec:intro}

The application of topology in the well established quantum band theory enabled material scientists to explore a wide array of novel phases of matter. The first and most widely studied are the topological insulators (TI), initially discovered in graphene sheets \cite{novoselov2004,novoselov2DGas2005,kaneQuantumSpinHall2005} but soon predicted in other two- and three-dimensional materials \cite{fu2007,wan2011,hasanReviewTopologicalInsulators2010,armitage2018,tokura2019}. TIs have the distinct property of hosting conducting edge states while the bulk remains electrically insulating, promising direct applications ranging from quantum information to spintronics \cite{hasanReviewTopologicalInsulators2010,he2019,fu2008,wang2017}.

Among the first 3D TIs found, the \BiSe and related family of materials quickly became the prototypical platform for exploring the properties of Dirac fermions, as only a single Dirac cone is responsible for the topological surface states \cite{zhang2009}. \BiSe gained special attention due to a large direct bulk band gap of over $200$ meV \cite{martinez2017}, allowing relativistic quasiparticles to persist even at room temperature \cite{hsieh2009bi2se3}. However, further applications of topological phenomena have been limited, as bulk electrons originating from structural defects shift the Fermi level of \BiSe into the conducting band and completely dominate the surface states in most transport experiments.

In fact, this family has long been known as degenerate semiconductors, and Bi$_2$Te$_3$ specifically can be readily made $n-$ or $p-$type depending on the defect favoured through the growth procedure \cite{zhang2013,wang2011,satterthwaite1957}. This, in fact, has made Bi$_2$Te$_3$ the parent compound on the majority of currently used Peltier devices for near room-temperature cooling \cite{beretta2019a}. \BiSe, on the other hand, has Se vacancies (\subt{V}{Se}) as the main defect present in stoichiometric samples, unavoidably donating electrons into the system \cite{hyde1974,navratil2004,hor2009} and making a $p-$type \BiSe hard to obtain.

As a consequence, \BiSe received much less attention until recent years, when the prediction of a 3D TI phase brought interest into suppressing these bulk charge carriers and tuning the Fermi level into the bulk band gap and topological Dirac band crossing. It is known that the charge carrier density can be reduced to $10^{17}$ cm$^{-3}$ by increasing the Se concentration during growth \cite{kohler1973,hyde1974,kohler1975,butch2010,analytis2010,dai2016}, and it can be further reduced to $10^{16}$ cm$^{-3}$ by Sb-doping the Se-rich samples \cite{analytis2010Nature}. Recently, it has been shown that the Fermi level can be lowered further to yield a $p$-type or nonconducting \BiSe by doping it with Ca \cite{hor2009,checkelsky2009}. In such samples, however, the mean free path seems to be reduced by an order of magnitude \cite{hor2009}, and additional bulk in-gap states of unknown origin become apparent \cite{checkelsky2009}. This suggests that a chemically simpler $p$-type \BiSe is desirable to resolve this effect, either as an intrinsic property of the material or originated solely due to extrinsic doping defects.

The coexistence of a wide array of defects in \BiSe adds an extra layer of complexity to the origin of the bulk conductivity. In stoichiometric and Se-rich samples, STM measurements have shown that the density of electrons can be largely related to the density of \subt{V}{Se} \cite{alpichshev2012}. Theoretical calculations have supported \subt{V}{Se} as the structural defect with lowest formation energy in a wide range of Bi:Se proportions \cite{zhang2013,west2012,xue2013,wang2012}, but rather surprisingly, Zhang \etal predicted that a $p$-type \BiSe might be obtained in extremely Bi-rich growth conditions, where the acceptor-type substitution of Bi in the Se site (\subt{Bi}{Se}) becomes the most stable defect \cite{zhang2013}. \subt{Bi}{Se} substitution has in fact been reported in samples grown with slight Bi excess \cite{urazhdin2002,dai2016}, but due to band-bending effects near the surface of \BiSe, the bulk energy states associated to each defect cannot be determined \cite{urazhdin2002,analytis2010}. More so, with the maximum reported Bi proportion of 45 \% \cite{urazhdin2002} in the initial growth mixture, the prediction of a $p$-type \BiSe in the Bi-rich portion of the phase diagram remains unrealized.

In this work, we performed a systematic study of \BiSe on single-crystalline samples grown in the entire Bi:Se phase diagram range available for the self-flux method. Following structural characterization through powder X-ray diffraction (PXRD) and energy dispersive spectroscopy (EDS), the electronic transport properties were determined through measurements of the resistivity, Hall effect and Shubnikov-de Haas oscillations, allowing insights on the defect formations. With recent works taken into consideration, our results provide a cohesive picture of the bulk properties of \BiSe, addressing previously overlooked issues in literature like the possibility of a Bi-rich $p$-type \BiSe \cite{zhang2013} and a topological phase transition proposed to explain a nonzero Berry phase in the bulk band \cite{kumar2015}.

\section{Experimental Methods}\label{sec:exp}

High-quality single crystals of Bi$_2$Se$_3$ were obtained through the self-flux method.  High-purity elemental Bi (Alfa-Aesar, 99.999\%) and Se (Sigma-Aldrich, 99.999\%) were loaded in evacuated quartz ampoules, kept at \SI{750}{\celsius} for $12$~h and then slowly cooled from \SI{710}{\celsius} down to \SI{630}{\celsius} at a \SI{1}{\celsius/h} rate, at which point the ampoules were centrifuged to separate excess flux from the grown crystals. Being a congruently melting phase, the compound can be obtained through a Bi-Se initial mixture with concentrations ranging from 49 \% up to 72 \% of Se \cite{ASM}, so $10$ different batches were prepared with concentrations evenly spaced between 52\% and 70\% of Se. Plate-like single crystals with dimensions as large as $10\times8\times2$~mm$^{3}$ were systematically obtained, particularly in Bi-rich growths (lower Se \%). Hereafter, samples will be referred by the nominal Se proportion used in the initial mixture, e.g., a sample grown from the stoichiometric Bi:Se proportion of 40:60 is denoted as Se60.

The structural details of the obtained phase in each growth were analyzed through powder X-ray diffraction (PXRD), using a Stoe STADI-P diffractometer with Cu-K$_{\alpha 1}$ radiation. Elemental analysis was performed through energy-dispersive X-ray spectroscopy (EDS) on freshly cleaved surfaces of the single crystals, using the EDS module of a Compact JEOL JSM-6010LA Scanning Electron Microscope (SEM).

Bi$_2$Se$_3$ is reported to show slightly different carrier densities in samples from the same batch \cite{wu2016}, so the $5$-probe method was used to simultaneously measure longitudinal resistivity ($\rho_{xx}$), magnetoresistance (MR) and transverse resistivity ($\rho_{xy}$) on the same sample. The measurements were performed in a $9$~T EverCool-II Physical Property Measurement System (PPMS) from Quantum Design, on $2$ single-crystals from each batch with the magnetic field parallel to the $c$-axis. All data obtained under a magnetic field was symmetrized by removing odd (even) components from the raw MR ($\rho_{xy}$) signal, and the results at $2.5$~K were checked to match the standard approach of relating values at positive and negative magnetic fields \cite{pippard1989magnetoresistance}. In the analysis of the Shubnikov-de Haas effect, \emph{ad-hoc} weights were attributed to the data to ensure proper background subtraction \cite{analytis2010Nature}.

\section{Experimental Results}\label{sec:results}

\subsection{Crystal Structure}\label{ssec:structure}

Bi$_2$Se$_3$ and the related family of materials (Bi$_2$Te$_3$, Sb$_2$Se$_3$, etc) crystallizes in a rhombohedral primitive cell with centrosymmetric $R\bar{3}m$ space group \cite{zhang2013}. Planes formed by the individual atoms stack on top of each other in the $z$-direction, forming a so-called quintuple layer (QL) of Se-Bi-Se-Bi-Se atoms. Two adjacent QLs are bounded by weak van der Waals forces between the adjacent Se planes, resulting in plane-like single crystals where the $xy$-planes are easily exposed through exfoliation of the weakly bonded QLs.

\begin{figure*}[htb]
    \centering
    \includegraphics[width=\textwidth]{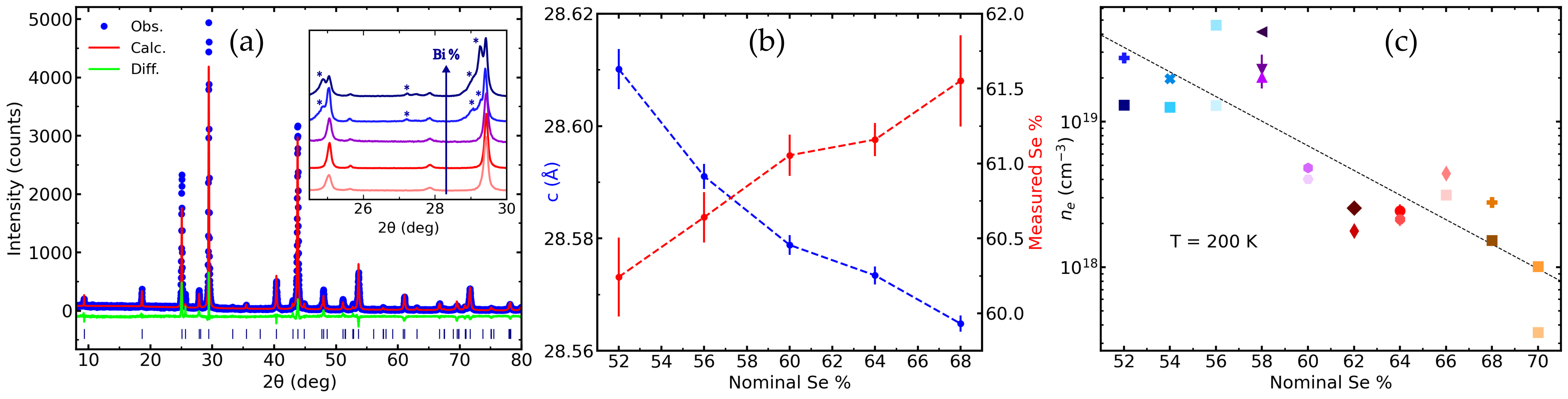}  
    \caption{\label{fig:1}(a) Powder X-ray diffraction of the Se64 \BiSe sample, with the measured data points, the calculated pattern and their difference. The inset shows the normalized diffraction in a shorter angle range for Se68, Se64, Se60, Se56 and Se52, where additional peaks related to the (Bi$_2$Se$_3$)$_x$(Bi$_2$)$_y$ family are marked with *. (b) In blue, the crystallographic $c$-axis as a function of initial Se concentration at growth, relating the decreasing QL distance with Se \%. In red, the effective Se compositions in the grown crystals, measured with EDS. (c) Density of electrons measured through the Hall effect at $200$ K as a function of Se concentration at growth. Dotted line is an exponential trend line. The colour code used in this article is shown here, with orange for Se-rich samples, blue for Bi-rich and red for close to stoichiometric. }
\end{figure*}

Rietveld refinement of the powder X-Ray diffraction patterns (Fig. \ref{fig:1}a) confirmed the expected crystalline structure of the obtained phase. For the stoichiometric and Se rich mixtures, Bi$_2$Se$_3$ is the only detected phase with a refinement $\chi^2$ reaching as low as 2.00. The inset of Fig. \ref{fig:1}a shows, however, extra peaks in Bi-rich samples, confirming the presence of secondary phases of the (Bi$_2$Se$_3$)$_x$(Bi$_2$)$_y$ family of materials, formed by periodically inserted Bi$_2$ layers between the QLs \cite{goncalves2017,goncalves2018}. 

Figure \ref{fig:1}b shows the $c$ unit cell parameter decreasing with an increasing Se proportion in the mixture. Since the $c$ parameter is directly dependent on the interplanar distance between QLs, this monotonic increase of the QL distance in Bi rich mixtures is a direct consequence of the different defects intrinsically present in Bi$_2$Se$_3$, as will be discussed in Section \ref{sec:discussion}. 

Conversely, the effective Se concentrations as measured by EDS (Fig. \ref{fig:1}b), grows with the nominal Se proportion, as expected. Since the majority of the pertinent defects (Se vacancies V$_\text{Se}$, Bi substitution on the Se site Bi$_\text{Se}$ and interstitial Bi \cite{dai2016,zhang2013,alpichshev2012,urazhdin2002}) would reduce the concentration of Se from stoichiometry, the effective values above 60 \% are attributed to the limited accuracy of the EDS experiments in determining absolute concentration values, only the trends. An alternative explanation, like the presence of interstitial Se atoms between the QLs \cite{dai2016}, seems unlikely considering the clear signatures of Bi interstitial atoms in our XRD results.

\subsection{Charge Carrier Properties}\label{ssec:carriers}

The negative and constant slope observed in the transverse resistivity $\rho_{xy}$ under an applied magnetic field confirms that all our single-crystals have $n$-type conductivity dominated by a single band of electrons, implying that an undoped $p$-type Bi$_2$Se$_3$ grown through the flux method remains elusive. \includecomment{Should we show a plot of these? Does it justify having a supplementary material?} 

The Hall coefficient $R_H$ was determined from the linear fit of the low-field $\rho_{xy}$ and, by considering the relation $R_H = r / (- e n_e)$, where $-e$ is the electron charge and $r$ is the Hall scattering factor (taken as unity), the electron density $n_e$ was obtained. As expected for an extrinsic semiconductor whose conductance comes mainly from doping defects, the charge carrier density does not depend significantly on temperature, but it can be tuned by the elemental proportion of the initial growth mixture. Figure \ref{fig:1}c shows that the electron density ranges from $10^{17}$~cm$^{-3}$ up to $10^{19}$~cm$^{-3}$ in Se and Bi-rich mixtures, respectively, consistent with previous reports \cite{kohler1973,hyde1974,butch2010,analytis2010}.

\begin{figure*}
    \centering
    \includegraphics[width=\textwidth]{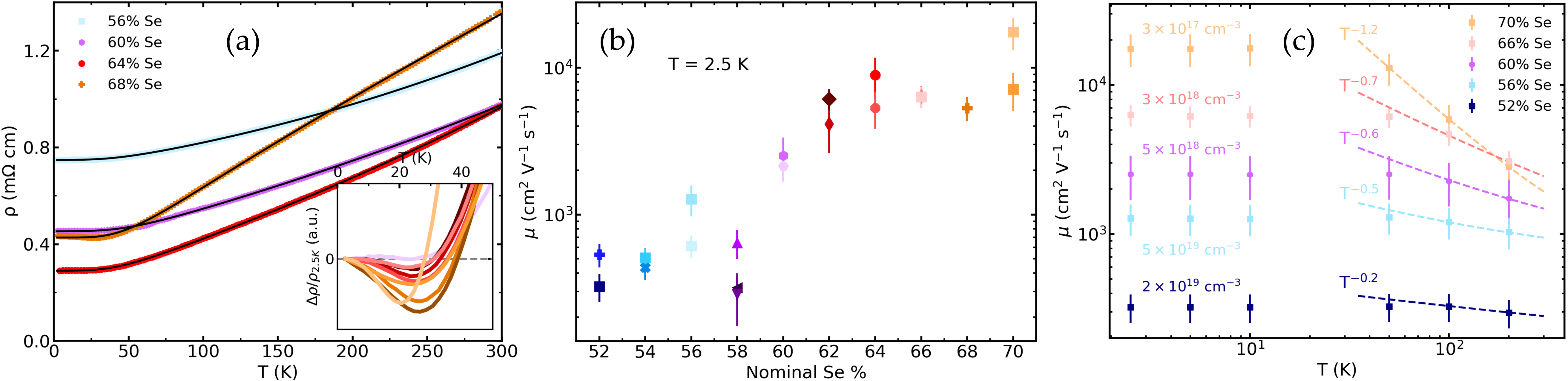}
    \caption{\label{fig:2} (a) Resistivity as a function of temperature for $4$ representative \BiSe crystals (Se56, Se60, Se64 and Se68). Black lines are fits to Eq. \ref{eq:BG}. The inset displays the normalized increase in resistivity below $30$ K for samples grown with high Se concentration. (b) Mobility at $2.5$ K as a function of initial Se \% used at growth. (c) Mobility as a function of temperature for \BiSe.  The carrier density $n_e$ of each sample is displayed, showing the transition from typical semiconductor behaviour in LCC to a temperature independent metallic behaviour at HCC.}
\end{figure*}

The longitudinal resistivity $\rho$ shows typical metallic behaviour, with a linear decrease upon cooling from high temperatures until freezing below $30$ K at a residual resistivity of order $10^{-1}$~m$\Omega$cm (Fig. \ref{fig:2}a). This trend is typical of a dominating electron-phonon scattering, that can be quantified with a Bloch-Grüneisen (BG) model \cite{ziman2001electrons} considering an additional $T^2$ dependence that accounts for the Fermi liquid electron-electron scattering \cite{ziman2001electrons,cao2012,vandermarel2011}
\begin{equation}\label{eq:BG}
\begin{split}
        \rho =& \rho_0 + B \qty(\frac{T}{\theta_{BG}})^2 +\\
             &+ A \qty(\frac{T}{\theta_{BG}})^{5} \int^{\theta_{BG}/T}_0 \frac{x^{5}}{(e^x - 1)(1  - e^{-x})} dx ,
\end{split}
\end{equation}
where $\rho_0$ is the residual resistivity and $\theta_{BG}$ is the BG temperature, related to the Debye temperature \cite{ziman2001electrons}. $A$ and $B$ are fitting parameters scaling the electron-phonon and electron-electron scattering mechanisms, respectively. Low carrier concentration samples (LCCs)  present two features that will be further discussed in Section \ref{sec:discussion} but are worth mentioning: (i) the resistivity below $\sim30$~K shows an upturn that has been previously attributed to impurity bands (inset Fig. \ref{fig:2}a) \cite{analytis2010,kohler1975,wiedmann2016}, but we will propose a simpler mechanism based on an increase in mobility; (ii) the electron-phonon coupling (parameter $A$) is greatly enhanced (Fig. \ref{fig:4}b) as a consequence of the reduced defect density.

The Hall mobility $\mu$ of the charge carriers at a given temperature is defined as $\mu = |R_H|/\rho$. Figure \ref{fig:2}b shows the calculated values at $T=2.5$~K, where low electron mobilities of order $10^2$~cm$^{2}$/V s are found in high carrier concentration samples (HCCs) and high mobilities of order $10^4$~cm$^{2}$/V s are seen on LCCs. As this is obtained at low temperatures without active phonons, it is a clear indication of increased scattering from structural defects in HCCs, supporting the idea that the defect density increases at lower Se \% and that these are responsible for donating charge carriers as well as serving as scattering centres.

Figure \ref{fig:2}c shows the temperature dependence of the mobility for five representative samples. With the number of data points taken, the mobility seems to be constant at low temperatures but starts decaying at $T > 50$~K, with the temperature dependence going from $\mu \propto T^{-1.2}$ for LCCs (high Se \%) down to $\mu \propto T^{-0.2}$ on HCCs (low Se \%). This is similar to previously reported results \cite{navratil2004,kohler1975,gao2014}, but samples with charge carrier densities on the order of $10^{16}$ cm$^{-3}$ show an increase in the mobility at low temperatures, attributed to ionized impurity scattering \cite{kohler1975}. For this reason, theoretical models for \BiSe have been developed in the past that consider scattering from ionized impurities dominating at low temperatures and acoustic phonons at high temperatures \cite{navratil2004}. In standard semiconductors, these would result in $\mu \propto T^{+3/2}$ and $\mu \propto T^{-3/2}$ respectively \cite{ziman2001electrons}, but deviations from these values into a temperature independent mobility are expected in highly degenerate semiconductors and metals \cite{niedermeier2017,cetnar2019}. Although our results are consistent with this picture at high temperatures, with the exponent approaching the expected $T^{+3/2}$ dependence for LCC crystals, the constant mobility at low temperatures across all samples may indicate that the scattering is not solely from charged defects, but also from charge-neutral interstitial impurities that contribute to temperature independent behaviour \cite{erginsoy1950,brooks1955,lu2019b,jansen2013}.

\subsection{Shubnikov-de Haas oscillations}\label{ssec:sdh}

\begin{figure*}[htb]
    \centering
    \includegraphics[width=\textwidth]{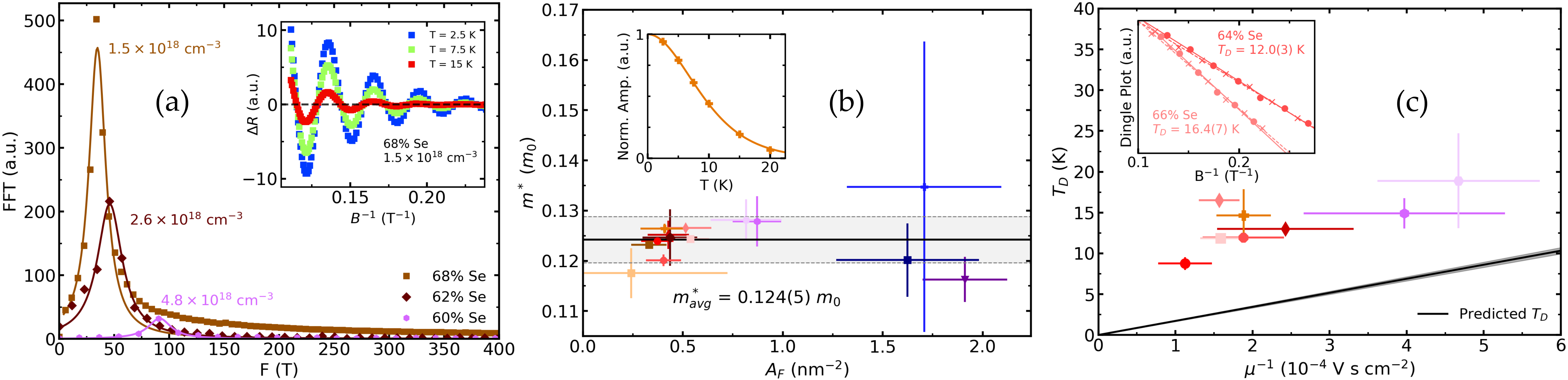}
    \caption{\label{fig:3} (a) Fast Fourier Transform of the Shubnikov-de Haas oscillations for \BiSe crystals obtained with different Se\% at growth at $2.5$ K. The inset shows the periodic oscillations in resistance as a function of inverse magnetic field at different temperatures, for a Se68 sample. (b) Effective masses as a function of Fermi surface area perpendicular to $B$, calculated through the temperature dependence of the oscillations. The inset shows the fit of the peak to trough amplitude as a function of temperature for a Se68 sample. (c) Dingle temperature determined from the SdH effect as a function of the inverse Hall mobility. The solid line represents the expected trend, showing a clear offset. The inset shows the Dingle plot for Se64 and Se66 samples.}
\end{figure*}

After subtracting a background from the raw magnetoresistance (MR), oscillations due to the Shubnikov-de Haas effect (SdH) are clearly revealed in the low temperature resistivity of LCCs. SdH oscillations could not be resolved in HCCs since magnetic fields above $B=9$ T are required, due to the reduced carrier mobility \cite{kulbachinskii1999}. The amplitude of the oscillations decays rapidly with increasing temperature and exponentially with $B^{-1}$, as shown in the inset of Fig. \ref{fig:3}a for a Se68 sample with carrier density of $10^{18}$ cm$^{-3}$.
The frequency is obtained by fitting a peak function to the Fast Fourier Transform (FFT) of the signal (Fig. \ref{fig:3}a), where both the frequency and amplitude of the oscillations are significantly sample dependent. The uncertainty is taken as the half width at half maximum. Theoretically, the frequency of oscillation $F$ is determined by the Onsager relation $F = (\hbar / 2 \pi e) A_F$, where $\hbar$ is the reduced Planck constant and $A_F$ is the cross-sectional area of the Fermi surface perpendicular to $B$ (in this case, orthogonal to the $k_z$ direction of the Brillouin Zone). In Section \ref{sec:discussion} a model will be constructed for the Fermi surface, enabling the relation between the frequency $F$ and the electron density $n_e$.

The overall shape of the SdH oscillation in the resistance $R$ is given by \cite{willardson1967} 
\begin{equation}\label{eq:SdH}
    \frac{\Delta R(B)}{R(0)} = \qty(\frac{B}{F})^{1/2} R_T R_D R_S \text{cos}\qty[2 \pi \qty(\frac{F}{B} - \phi)  ],
\end{equation}
where $R_T$, $R_D$ and $R_S$ are respectively the temperature, Dingle and spin damping factors. The phase of oscillation $\phi$ is given by $\phi = 1/2 - \beta/2\pi - \delta$, where $\delta=-1/8, +1/8, 0$ for 3D electrons, 3D holes or 2D carriers, respectively. The Berry phase $\beta$ is $\pi$ when the Fermi surface encloses a Dirac crossing in a topological material, but is null for trivial bands \cite{mikitik1999,taskin2011}.

The temperature dependence of the SdH oscillations is determined by the temperature damping factor $R_T$, defined as $R_T= \frac{\xi(B, T)}{\text{sinh}\qty[\xi(B, T)]}$, where $\xi(B, T) = 2 \pi^2 k_B  T m^* / (\hbar e B) $, $k_B$ is the Boltzmann constant and $m^*$ is the charge carrier's effective mass. By fitting $R_T$ to the amplitude peaks for different temperatures (inset Fig. \ref{fig:3}b), the effective mass $m^*$ of all samples average to $m^* = 0.124(5) m_0$, with $m_0$ as the free electron's rest mass.

The Dingle factor $R_D$  describes the damping of the oscillations from the scattering of the charge carriers. It is defined as $R_D = \text{exp}\qty[-\xi(B, T_D)]$, where the Dingle temperature $T_D$ is inversely proportional to the finite carrier lifetime $\tau$ as $T_D = \hbar / (2 \pi k_B \tau)$. Considering the lower mobilities in HCC Bi$_2$Se$_3$ samples, lower $\tau$ explains why these samples have SdH oscillations that are harder to resolve. Considering the oscillations at a fixed temperature, the maxima and minima of Eq. \ref{eq:SdH} can be linearized  (inset Fig. \ref{fig:3}c) to find the Dingle temperatures, which result between $8$ K and $25$ K, consistent with previous reports between $7$ K and $30$ K \cite{mukhopadhyay2015, kohler1973,eto2010}.

The scattering time $\tau$ is defined as $\tau=m^* \mu/e$, so $T_D$ is predicted to be proportional to $\mu^{-1}$ (solid line in Fig. \ref{fig:3}c). The experimental data points do seem to follow this trend, but the Dingle temperatures are much larger than the predicted values. For these $T_D$ values, the mobility of the charge carriers in the Landau Levels (LLs) is up to six times smaller than those measured through the Hall effect. This reduced mobility has previously been reported for Bi$_2$Se$_3$ \cite{eto2010} as well as for many semiconductors \cite{shoenberg2009magnetic} and can be attributed to the difference in scattering times governing transport phenomena (e.g. Hall effect) and phase decoherence in quantum oscillations such as SdH \cite{eto2010,dassarma1985}. When small angle scattering is dominant, as is generally the case at low temperatures \cite{eto2010}, the loss of coherence happens faster. Furthermore, this difference has been theoretically predicted to increase exponentially in 2D materials with an additional scattering layer separated from the conducting layers \cite{dassarma1985}. In Bi$_2$Se$_3$ this scenario points to the significance of charge scattering from the interstitial impurities found between the conducting QLs and will be further discussed in Sec. \ref{sec:discussion}.

We now address the phase of the SdH oscillations by constructing the Landau Fan Diagram (LFD), where LL indices $n$ are assigned to each maxima of the SdH oscillations and plotted as a function of the inverse field values in which they occur. The choice of using the maxima \cite{eto2010,xiong2012,murakawa2013,he2014,zhang2017} instead of the minima \cite{kumar2015,qu2010a,narayanan2015,park2011} of $\rho$ to construct the LFD was made based on recent theoretical developments, where both 2D \cite{taskin2011} and 3D \cite{wang2016} structures have been shown to present a maximum in $\rho$ (or a sharp peak in the Quantum Hall Effect \cite{hasanReviewTopologicalInsulators2010}) when LLs crosses the Fermi level. 

Figure \ref{fig:4}a shows the LFDs for the samples that presented clear SdH oscillations, where the intercept of the solid lines with the ordinary axis is the phase $-\phi$. It is important to consider the Zeeman splitting through the spin damping factor $R_S$, defined as $R_S = \text{cos}\qty(\frac{\pi}{2} g \mu^*)$ where $g$ is the electron spin $g$-factor, since $R_S < 0$ would change $\phi$ by $\frac{\pi}{2}$ and render the determination of the topological Berry phase useless. Fortunately, several works have shown that for the main bulk band of Bi$_2$Se$_3$, the Zeeman splitting energy $g \mu_B B$, where $\mu_B=e\hbar/2m_0$ is the Bohr magneton, is precisely double the LL splitting energy $\hbar e B / m^*$ \cite{orlita2015,kohler1975g-factor,fauque2013,mukhopadhyay2015}. This fixes $g=4 m_0 / m^*$ and results in $R_S=1$.

Considering that the main band is composed of 3D electrons, we can take $\delta=-1/8$ and obtain the Berry phase (inset Fig. \ref{fig:4}a). The results agree with the majority of reports in literature \cite{chen2022,busch2018}, where $\beta$ approaches $0$ and the main conductive band is topologically trivial in all samples. A recent contradicting claim of a topological semimetal band, made after an obtained Berry phase of $\beta = \pi$ \cite{kumar2015}, can be questioned since it was determined by assigning the minima of the oscillations to the LL crossing in the LFD, which changes the calculated Berry phase by $\pi$.

\section{Discussion}\label{sec:discussion}

\begin{figure*}[htb]
    \centering
    \includegraphics[width=\textwidth]{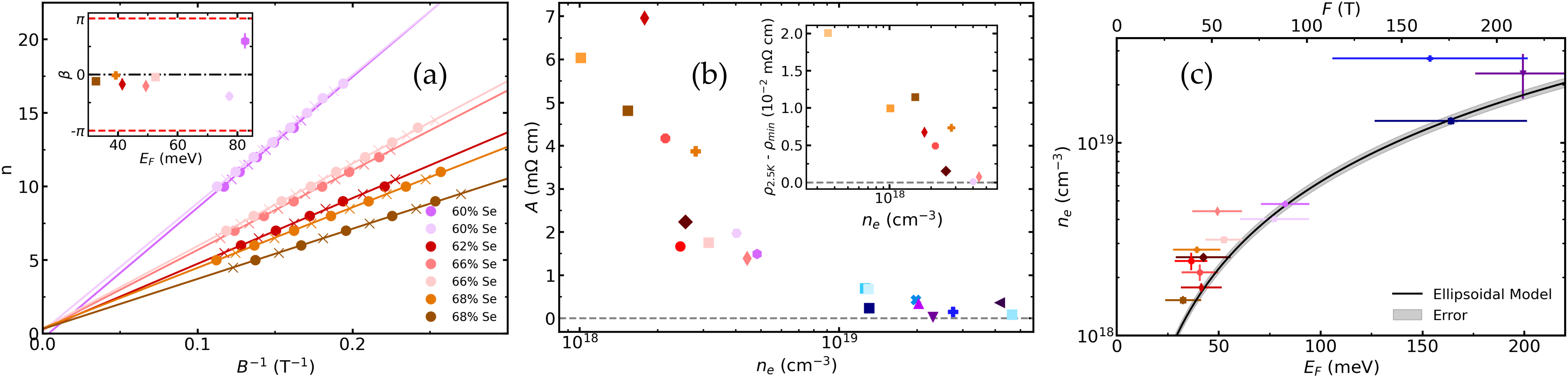}
    \caption{\label{fig:4}(a) Landau fan diagram for the different \BiSe samples. The inset shows the calculated Berry phase as a function of the Fermi level, where it is shown to be close to $0$ for all samples. (b) The $A$ parameter of Eq. \ref{eq:BG} as a function of electron density, showing the decreasing electron-phonon interaction for higher electron densities. The inset displays the magnitude of the resistivity upturn, increasing for lower density samples. (c) Electron density measured through the Hall effect as a function of the Fermi level obtained through the SdH. The solid line shows the prediction of Eq. \ref{eq:dos}, evidencing that most electrons come from the main conduction band.}
\end{figure*}

The increase in the carrier density from $10^{17}$ cm$^{-3}$ to $10^{19}$ cm$^{-3}$ (Fig. \ref{fig:1}c) for lower Se (higher Bi) concentrations is consistent with the picture of charged Se vacancies (\subt{V}{Se}) being the main source of conducting electrons in \BiSe \cite{hyde1974,navratil2004,hor2009,dai2016,alpichshev2012}. The reduced mobility (Fig. \ref{fig:2}b) can also be attributed to enhanced scattering from a larger density of structural defects in Bi rich growths, but the nature of these defects cannot be solely attributed to \subt{V}{Se}. The extra peaks in the PXRD (inset Fig. \ref{fig:1}a) support the presence of interstitial Bi layers, and the increase in the interlayer distance (left axis in Fig. \ref{fig:1}b) is also predicted from other defects such as Bi substitution on the Se site \subt{Bi}{Se} \cite{zhang2013,dai2016,urazhdin2002} and Se on the Bi site \subt{Se}{Bi} \cite{zhang2013}.

The presence of multiple types of defects is further supported by the scattering processes of the charge carriers, manifested in the temperature dependence of transport experiments. A temperature independent Hall mobility (Fig. \ref{fig:2}c) is expected at low temperatures when charge carriers are scattered on electrically neutral defects \cite{erginsoy1950,brooks1955,lu2019b,jansen2013} and, since this is found even in LCCs samples (rich in Se), it cannot not be solely attributed to the degeneracy of the system. Furthermore, the significantly reduced lifetime of the electrons generating the SdH oscillations, evidenced by the larger than expected Dingle temperatures $T_D$ (Fig. \ref{fig:3}c) is an indication that defects outside the conducting layer are responsible for scattering \cite{dassarma1985}. These results point to the significance of interstitial atoms in transport phenomena for both Bi and Se-rich \BiSe samples, signalling the presence of not only the interstitial Bi layers found in the PXRD, but also of interstitial Se atoms in Se-rich samples, as previously reported in flux-grown single crystals \cite{dai2016}.

The absence of a $p$-type single crystal even in very Bi-rich initial mixtures shows that the \subt{Bi}{Se} substitution either appears in very low concentration (unable to compensate for the high density of \subt{V}{Se} defects) or it is not an acceptor-like defect as was predicted by Zhang \cite{zhang2013}. Note that \subt{Bi}{Se} as a donor-like substitution has been suggested in other theoretical works \cite{wang2012, west2012}.

The structural defects also play an important role on the propagation of phonons, as found in the temperature dependence of the resistivity $\rho$ through the parameter $A$ of the BG equation (Eq. \ref{eq:BG}). Being a measure of the electron-phonon (e-ph) scattering, the decrease of $A$ in HCCs can be attributed to the larger amount of structural defects effectively reducing the distance that lattice distortions can propagate coherently. Considering that the e-ph coupling increases with this distance \cite{gunst2016}, this shows that the e-ph coupling can be controlled through the density of structural defects, or experimentally through the growth procedure.

We now address the upturn in resistivity seen in Fig. \ref{fig:2}a for LCC \BiSe. This effect has previously been attributed to an impurity band with thermally activated charge carriers, where its conductivity only becomes apparent when electrons from the main bulk band "freeze" at low temperatures \cite{analytis2010,kohler1975,wiedmann2016}. In-gap impurity states have in fact been reported in thin films \cite{qu2013,bianchi2010}, and in carefully doped samples with very low carrier concentration (10$^{16}$ cm$^{-3}$) \cite{checkelsky2009}, but the linear Hall resistivity $\rho_{xy}$ found means that the mobility of these impurity carriers has to be considerably lower than the ones from the main band. Additionally, since the electron density seems temperature independent and the exponentially decaying $\rho$, expected for thermally activated carriers \cite{seeger2013semiconductor}, is absent, we propose an alternate simpler explanation based on the effects of ionized impurity scattering.

Following the discussion in Sec. \ref{ssec:carriers}, the scattering at low temperatures is likely a combination of neutral and ionized impurities, and these effects are expected to become more apparent in LCCs. A non-degenerate semiconductor dominated by ionized defect scattering presents a mobility that increases with temperature when there are no active phonons, and with a constant charge carrier density, the resistivity would decrease with temperature (Fig. \ref{fig:2}a). Combined with the fact that the absolute upturn in $\rho$ decreases somewhat exponentially with charge density $n_e$ (inset Fig. \ref{fig:4}b), \BiSe can be regarded as an extrinsic semiconductor where the effects of charged defect scattering become apparent first through the resistivity upturn in low carrier density samples.

The Fermi level $E_F$ of the samples can be calculated directly from the cross sectional area of the Fermi surface $A_F$ obtained in the SdH oscillations. Considering the constant effective mass and the topologically trivial nature of the main conduction band found through the SdH oscillations, a parabolic dispersion can be considered in the $k_z=0$ plane, resulting in $E_F = \hbar^2 A_F / 2 \pi m^*$. The Fermi level of our samples (Fig. \ref{fig:4}c) lies between $25$ meV and $200$ meV from the bottom of the conduction band. From previous angle-dependent SdH oscillation studies \cite{kohler1973trigonal,kohler1973,eto2010}, it is known that the Fermi surface of \BiSe has the shape of an ellipsoid elongated in the $k_z$ direction of the Brillouin zone. Despite the parabolic dispersion in the $k_x-k_y$ plane, there is a Fermi level-dependent flattening of the FS in the $k_z$ direction at higher $E_F$ \cite{kohler1973} that can be accounted for by an anisotropy factor $\eta=A_{zF}/A_{F}$. Following Mukhopadhyay \etal \cite{mukhopadhyay2015}, $\eta$ can be fitted to $E_F$ through a power law as $\eta = 1 + a_\eta E_F^{b_\eta}$, where $a_\eta = 1.32(9)$ and $b_\eta = 0.27(3)$ is obtained using previous reports \cite{mukhopadhyay2015,kohler1973,kulbachinskii1999,eto2010}. From the density of states of a spin-degenerate 3D Fermi gas \cite{ashcroft1976solid}, the effective mass in the $k_z$ direction ($m_z^*$) must be evaluated before the electron density can be obtained. As a first approximation, a phenomenological ellipsoidal model can be considered as $m_z^* = m^* \eta^2$ \cite{mukhopadhyay2015,kohler1973}, resulting in the electron density for \BiSe as
\begin{equation}\label{eq:dos}
    n_e = \frac{1}{3 \pi^2} \qty(\frac{2 m^* E_F}{\hbar^2})^{3/2} \qty(1 + \frac{3 a_\eta E_F^{b_\eta}}{2 b_\eta + 3}),
\end{equation}
represented by the solid curve in Fig. \ref{fig:4}c, which shows fair agreement with the experimental data obtained from Hall effect. 

Note that by taking this phenomenological model into consideration, discrepancies previously seen between the experimental carrier concentration and the one predicted from the main bulk band have been largely reduced \cite{kohler1973,eto2010}. Yet, the model still seems to predict slightly lower electron densities than what is found through the Hall effect, and although this is a simplified model, the presence of very low mobility carriers not originated from the main conduction band cannot be discarded. Indeed, this could be attributed to temperature-independent electrons from extrinsic impurities, inherent to all \BiSe samples grown here, or a secondary conduction band of low mobility electrons, as has been shown to be populated above $40$ meV for Bridgman-grown samples doped with Sb \cite{kulbachinskii1999}.

\section{Conclusion}\label{sec:conclusion}

The bulk conducting states of the topological insulator \BiSe were studied by means of transport experiments and quantum oscillations. High-quality single crystals were obtained in the entire concentration range available in the phase diagram for flux growths, allowing the construction of a concise picture of all undoped \BiSe that can be grown through the flux method.

Despite theoretical predictions \cite{zhang2013}, electron donor-type defects, likely \subt{V}{Se}, dominate all samples and shift the Fermi level into the main conduction band. Other types of impurities, however, have a significant role on the scattering of the charge carriers, both directly, as can be seen in the low temperature mobility and scattering time, and indirectly through the reduction of the electron-phonon coupling at higher temperatures.

Careful analysis of the Shubnikov-de Haas effect confirmed the bulk electrons as topologically trivial, clarifying earlier discrepancies \cite{kumar2015}. A phenomenological model allowed the determination of the origin of the vast majority of these electrons from the main conduction band, but the presence of additional low mobility bands cannot be fully disregarded. Further theoretical investigations considering the complex scattering mechanisms are required.

As a thorough overview of transport properties of bulk \BiSe, this work helps to shed light on the complex interplay of the inherent impurities and provides a reference for experimental fine-tuning the bulk conducting states of this flagship topological insulator.

\section*{Acknowledgements}\label{sec:acknowledgements}

We are grateful for J. Munevar, L. Mendonça-Ferreira, A.R.V. Benvenho, S. K. Singh, G. Vasquez and C. A. Sonego for fruitful discussions. We would also like to thank K. K. F. Barbosa and F. F. Ferreira for the PXRD measurements, and the UFABC multiuser central facilities for the experimental support. This work was funded by Brazilian agencies FAPESP (grant number 2017/10581-1), CAPES and CNPq.

\bibliography{references}


\end{document}